\documentclass[12pt,preprint]{aastex}

\def\ltsima{$\; \buildrel < \over \sim \;$}
\def\gtsima{$\; \buildrel > \over \sim \;$}
\def\lsim{\lower.5ex\hbox{\ltsima}}
\def\gsim{\lower.5ex\hbox{\gtsima}}
\def\lapp{\ifmmode\stackrel{<}{_{\sim}}\else$\stackrel{<}{_{\sim}}$\fi}
\def\gapp{\ifmmode\stackrel{>}{_{\sim}}\else$\stackrel{<}{_{\sim}}$\fi}

\newdimen\minuswidth    
\setbox0=\hbox{$-$}
\minuswidth=\wd0
\catcode`@=\active
\def@{\kern\minuswidth}
\setbox0=\hbox{\rm0}
 
\shorttitle{Blue Stragglers in NGC~1904}
\shortauthors{Lanzoni et al.}
 
\begin{document} 
 
\title{A Panchromatic Study of the Globular Cluster NGC~1904. I: The Blue
Straggler Population
\footnote{Based on observations with the NASA/ESA HST, obtained at the Space
Telescope Science Institute, which is operated by AURA, Inc., under NASA
contract NAS5-26555. Also based on GALEX observations (program GI-056) and
WFI observations collected at the European Southern Observatory, La Silla,
Chile, within the observing programs 62.L-0354 and 64.L-0439.}  }

\author{
B. Lanzoni\altaffilmark{1,2},
N. Sanna\altaffilmark{3},
F.R. Ferraro\altaffilmark{1},
E. Valenti\altaffilmark{4},
G. Beccari\altaffilmark{2,5,6},
R.P. Schiavon\altaffilmark{7},
R.T. Rood\altaffilmark{7},
M. Mapelli\altaffilmark{8},
S. Sigurdsson\altaffilmark{9}
}
\affil{\altaffilmark{1} Dipartimento di Astronomia, Universit\`a degli Studi
di Bologna, via Ranzani 1, I--40127 Bologna, Italy}
\affil{\altaffilmark{2} INAF--Osservatorio Astronomico di Bologna, via
Ranzani 1, I--40127 Bologna, Italy}  
\affil{\altaffilmark{3} Dipartimento di Fisica, Universit\`a degli Studi di
Roma Tor Vergata, via della Ricerca Scientifica, 1, I--00133 Roma, Italy}
\affil{\altaffilmark{4} European Southern Observatory,
 Alonso de Cordova 3107, Vitacura, Santiago, Chile}
\affil{\altaffilmark{5} Dipartimento di Scienze della Comunicazione, 
Universit\`a degli Studi di Teramo, Italy}
\affil{\altaffilmark{6} INAF--Osservatorio Astronomico 
di Collurania, Via Mentore Maggini, I--64100 Teramo, Italy}
\affil{\altaffilmark{7} Astronomy Department, University of Virginia,
P.O. Box 400325, Charlottesville, VA, 22904}
\affil{\altaffilmark{8} University of Z\"urich, Institute for Theoretical
Physics, Winterthurerstrasse 190, CH-8057 Zurich}

\affil{\altaffilmark{9} Department of Astronomy and Astrophysics, The
Pennsylvania State University, 525 Davey Lab, University Park, PA~16802}

\date{30 March, 07}


\begin{abstract}
By combining high-resolution (HST-WFPC2) and wide-field ground based (2.2m
ESO-WFI) and space (GALEX) observations, we have collected a multi-wavelength
photometric data base (ranging from the far UV to the near infrared) of the
galactic globular cluster NGC1904 (M79). The sample covers the entire cluster
extension, from the very central regions up to the tidal radius.  In the
present paper such a data set is used to study the BSS population and its
radial distribution. A total number of 39 bright ($m_{218}\le 19.5$) BSS has
been detected, and they have been found to be highly segregated in the
cluster core. No significant upturn in the BSS frequency has been observed in
the outskirts of NGC~1904, in contrast to other clusters (M~3, 47~Tuc,
NGC~6752, M~5) studied with the same technique.  Such evidences, coupled with
the large radius of avoidance estimated for NGC~1904 ($r_{\rm avoid}\sim 30$
core radii), indicate that the vast majority of the cluster heavy stars
(binaries) has already sunk to the core.  Accordingly, extensive dynamical
simulations suggest that BSS formed by mass transfer activity in primordial
binaries evolving in isolation in the cluster outskirts represent only a
negligible (0--10\%) fraction of the overall population.
\end{abstract}

\section{INTRODUCTION}

Blue straggler stars (BSS) appear brighter and bluer than the Turn-Off (TO)
point along an extension of the Main Sequence in color-magnitude diagrams
(CMDs) of stellar populations.  Hence, they mimic a young stellar population,
with masses larger than the normal cluster stars \citep[this is also
confirmed by direct mass measurements; e.g.][]{sha97}. BSS are thought to be
objects that have increased their initial mass during their evolution, and
two main scenarios have been proposed for their formation
\citep[e.g.,][]{bai95}: the {\it collisional scenario} suggests that BSS are
the end-products of stellar mergers induced by collisions (COL-BSS), while in
the \emph{mass-transfer} scenario BSS form by the mass-transfer activity
between two companions in a binary system (MT-BSS), possibly up to the
complete coalescence of the two stars \citep{mat90, prit91, BaiPins95,
carn05, tian06, leigh07}. Hence, understanding the origin of BSS in stellar
clusters provides valuable insight both on the binary evolution processes and
on the effects of dynamical interactions on the (otherwise normal) stellar
evolution.  The MT formation scenario has by recently received further
support by high-resolution spectroscopic observations, which detected
anomalous Carbon and Oxygen abundances on the surface of a number of BSS in
47 Tuc \citep{COdep}. However the role and relative importance of the two
mechanisms are still largely unknown.

To clarify the BSS formation and evolution processes we studying the BSS
radial distribution over the entire cluster extension in a number of galactic
globular clusters (GCs).  We completed such studies in 5 GCs: M~3
\citep{fe97}, 47~Tuc \citep{fe04}, NGC~6752 \citep{sab04}, $\omega~$Cen
\citep{fe06b}, and M~5 \citep[][see also Warren, Sandquist \& Bolte
2006]{lan07}.  Apart from $\omega\,$Cen where mass segregation processes have
not yet played a major role in altering the initial BSS distribution, the BSS
are always highly concentrated in the cluster central regions. Moreover, in
M~3, 47~Tuc, NGC~6752, and M~5 the BSS fraction decreases at intermediate
radii and rises again in the outskirsts of the clusters, yielding a
\emph{bimodal} distribution. Preliminary evidences of such a bimodality have
been found also in M~55 by \citet{zag97}.
Recent dynamical simulations \citep{ma04,ma06,lan07} have been used to
interpret the observed trends and have shown that a significant fraction
($\gsim 50$\%) of COL-BSS is required to account for the observed BSS central
peaks. In addition, a fraction of 20-40\% MT-BSS is needed to reproduce the
outer increase observed in these clusters.  The case of $\omega~$Cen is
reproduced by assuming that the BSS population in this cluster is composed
entirely of MT-BSS. These results demonstrate that detailed studies of the
BSS radial distribution within GCs are very powerful tools for better
understanding the BSS formation channels and for probing the complex
interplay between dynamics and stellar evolution in dense stellar systems.

In this paper we present multi-wavelength observations of NGC~1904.  These
observations are part of a coordinated project aimed at properly characterize
the UV excess of old stellar aggregates as globular clusters, in terms of
their hot stellar populations, like Horizontal Branch (HB) and Extreme HB
stars, post-Asymptotic Giant Branch stars, BSS, etc. From integrated light
measurements obtained with UIT \citep[see][]{dorm95}, NGC~1904 was known to
be relatively bright in the UV, and it was selected as a prime target in both
our high-resolution (using HST) and wide-field (using GALEX) UV surveys.  We
have obtained a large set of data: {\it (i)} high-resolution ultraviolet (UV)
and optical images of the cluster center have been secured with the WFPC2 on
board HST; {\it (ii)} complementary wide-field observations covering the
entire cluster extension have been obtained in the UV and optical bands by
using the far- and near-UV detectors on board the {\it Galaxy Evolution
Explorer} (GALEX) satellite and with ESO-WFI mounted at the 2.2\,m ESO
telescope, respectively.  The combination of these datasets allowed a study
of the structural properties of NGC~1904 (thus leading to an accurate
redetermination of the center of gravity and the surface density profile),
and of the radial distribution of the evolved stellar populations (in
particular the BSS and horizontal branch star distributions have been derived
over the entire cluster extension).  While a companion paper (Schiavon et
al. 2007, in preparation) will focus on the morphology and the structure of
the HB, the present paper is devoted to the BSS population.

\section{OBSERVATIONS AND DATA ANALYSIS}
\subsection{The data sets}
\begin{figure}[!hp]
\begin{center}
\includegraphics[scale=0.7]{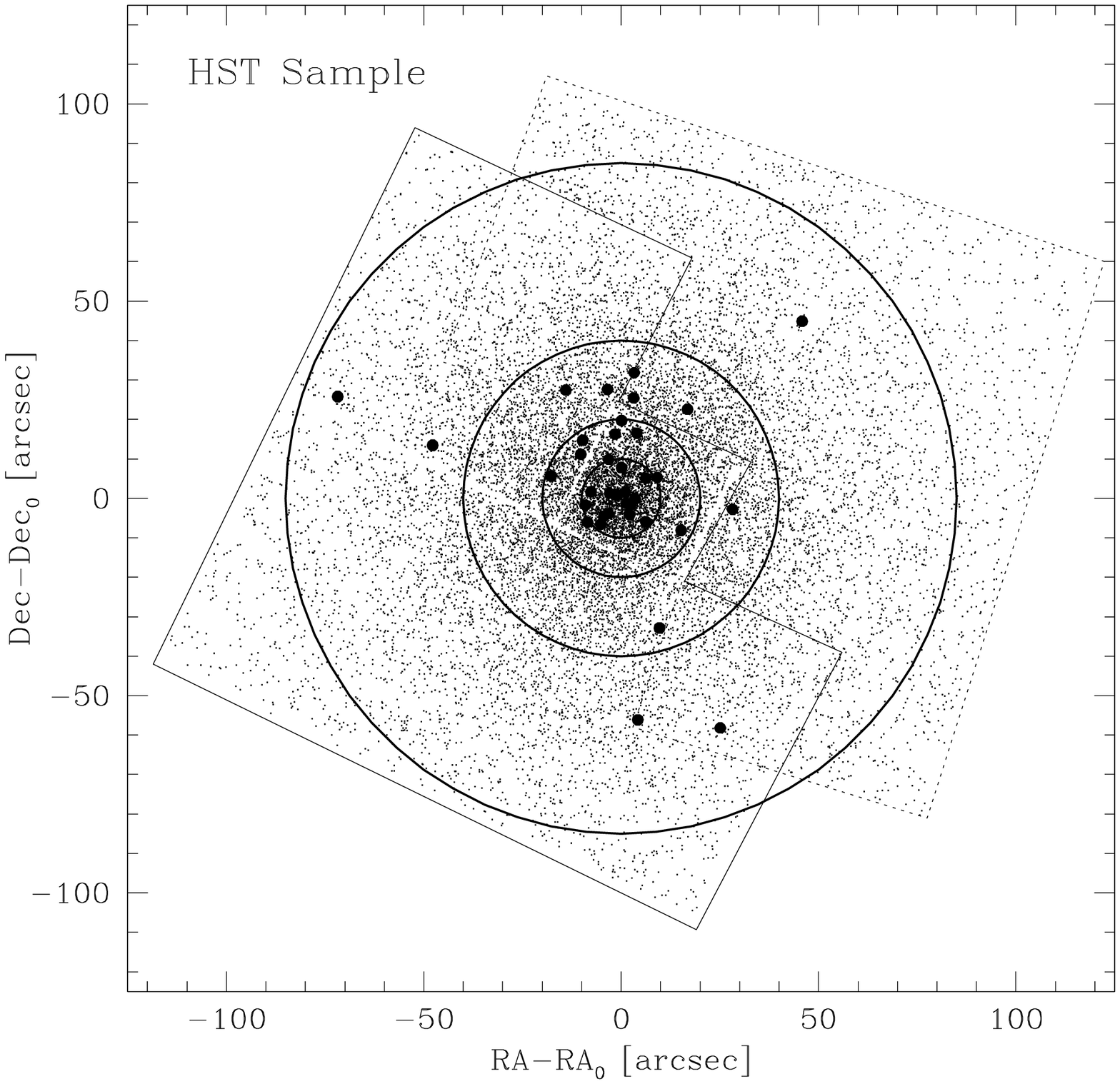}
\caption{Map of the combined HST sample.  The light solid and dotted lines
delimit the FoVs of Pointing B and A, respectively.  Star positions are
plotted with respect to the center of gravity $C_{\rm grav}$ derived in
Section \ref{sec:cgrav}: $\alpha({\rm J2000}) = 05^{\rm h}\, 24^{\rm m}\,
11.\fs09$, $\delta ({\rm J2000})= -24^{\rm o}\, 31\arcmin\, 29\farcs 00$.
The positions of all BSS identified in this sample are marked with heavy dots
and the concentric annuli used to study their radial distribution (cfr. Table
1) are also shown. The inner and outer annuli correspond to $r=r_c=10\arcsec$
and $r=85\arcsec$, respectively.}
\label{fig:HST}
\end{center}
\end{figure}

The present study is based on a combination of different photometric data
sets:

\emph{1. The high-resolution set} -- It consists of a series of UV, near UV
and optical images of the cluster center obtained with HST-WFPC2 with two
different pointings.  In both cases the Planetary Camera (PC, the highest
resolution instrument with $0\farcs 046\, {\rm pixel}^{-1}$) has been pointed
approximately on the cluster center to efficiently resolve the stars in the
highly crowded central regions; the three Wide Field Cameras (WFC with
resolution $0\farcs 1\, {\rm pixel}^{-1}$) have been used to sample the
surrounding regions.  Observations in {\it Pointing A} (Prop. 6607,
P.I. Ferraro) have been performed through filters F160BW (far-UV), F336W
(approximately an $U$ filter) and F555W ($V$), for a total exposure time
$t_{\rm exp}=3300$, 4400, and 300 sec, respectively.  {\it Pointing B} is a
set of public HST-WFPC2 observations (Prop.  6095, P.I. Djorgovski) obtained
through filters F218W (mid-UV), F439W ($B$) and F555W ($V$). Because of the
different orientations of the four cameras, this data set is complementary to
the former (with the PC field of view in common), thus offering full coverage
of the innermost regions of the cluster (see Figure \ref{fig:HST}).  The
combined photometric sample is ideal for efficiently studying both the hot
stellar populations (as the BSS and the HB stars) and the cool red giant
branch (RGB) population, and to guarantee a proper combination with the
wide-field data set (see below).

The photometric reduction of both the high-resolution sets was carried out
using ROMAFOT \citep{buon83}, a package developed to perform accurate
photometry in crowded fields and specifically optimized to handle
under-sampled Point Spread Functions \citep[PSFs;][]{buon89}, as in the case
of the HST-WFC chips.  The standard procedure described in
\citet[][2001]{fe97} was adopted to derive the instrumental magnitudes and to
calibrate them to the STMAG system by using the zero-points of
\citet{holtz95}. The magnitude lists were finally cross-correlated in order
to obtain a combined catalog.

\begin{center}
\begin{figure}[!p]
\includegraphics[scale=0.7]{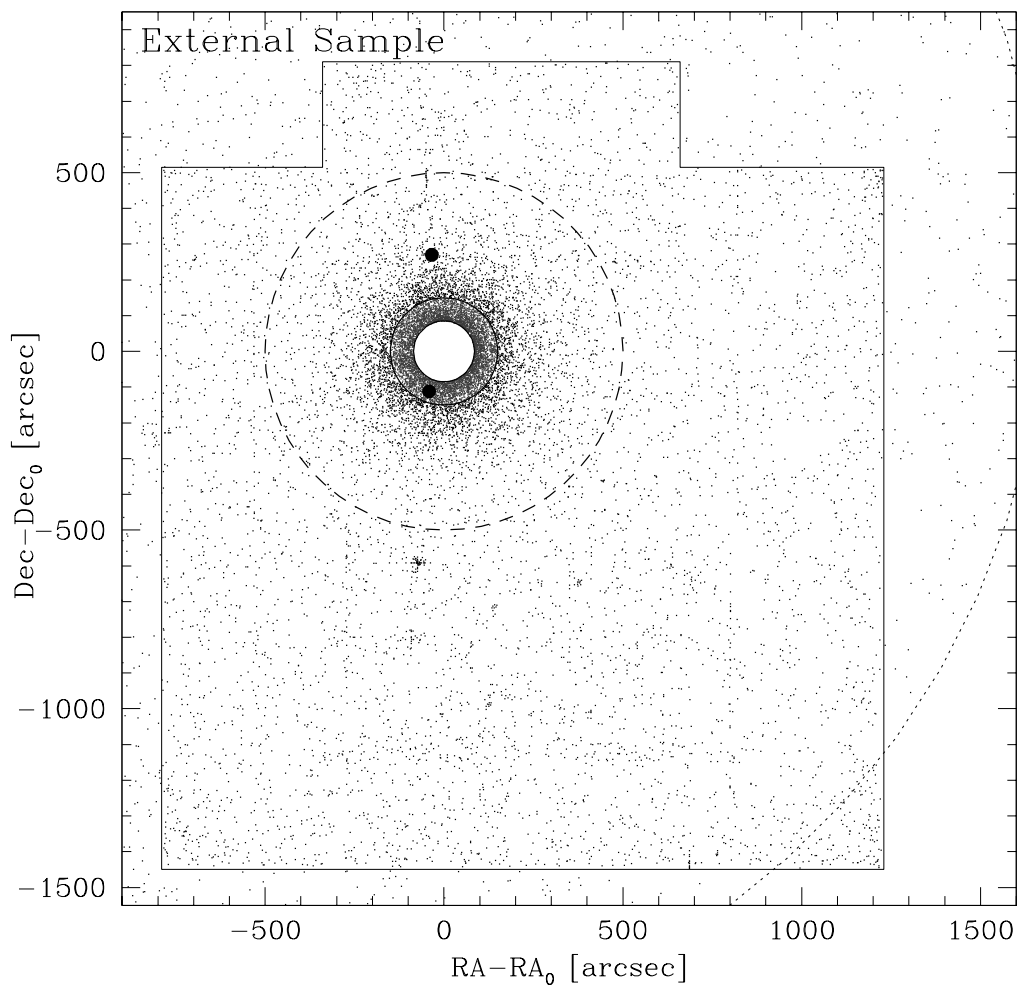}
\caption{Map of the External sample. The light solid and dotted lines delimit
the WFI and the GALEX FoVs, respectively. The two BSS detected in the
External sample are marked as heavy dots, and the concentric annuli used to
study their radial distribution are shown as heavy circles. The inner annulus
is at $85\arcsec$ and corresponds to the most external one in Figure
\ref{fig:HST}. The heavy dashed circle marks the tidal radius of the cluster
($r_t\simeq500\arcsec$).}
\label{fig:Ext}
\end{figure}
\end{center}

\emph{2. The wide-field set} - A complementary set of wide-field $U,~B,$ and
$I$ images was secured by using the Wide Field Imager (WFI) at the 2.2m
ESO-MPI telescope, during an observing run in January 1999 (Progr.~ID:
062.L-0354, PI: Ferraro). A set of WFI $V$ images (Progr.~ID: 064.L-0255) was
also retrieved from the ESO-STECF Science Archive.  Additional deep
wide-field images were obtained in the UV band with the satellite GALEX
(GI-056, P.I. Schiavon) through the FUV (1350--1750\,\AA) and NUV
(1750--2800\,\AA) detectors.  With a global field of view (FoV) of
$34\arcmin\times 34\arcmin$, the WFI observations cover the entire cluster
extension. There is also full coverage of the cluster in the UV thanks to the
large GALEX FoV, which is approximately $1\deg$ in diameter and includes the
WFI FoV (see Figure \ref{fig:Ext}, where the cluster is roughly centered on
WFI CCD $\# 2$). However, because of the low resolution of the instrument
($4\arcsec$ and $6\arcsec$ in the FUV and NUV channels, respectively), GALEX
data have been used to sample only the external cluster regions not covered
by HST.

The raw WFI images were corrected for bias and flat field, and the overscan
regions were trimmed using IRAF\footnote{IRAF is distributed by the National
Optical Astronomy Observatory, which is operated by the Association of
Universities for Research in Astronomy, Inc., under cooperative agreement
with the National Science Foundation.} tools ({\tt mscred} package). Standard
crowded field photometry, including PSF modeling, was carried out
independently on each image using DAOPHOTII/ALLSTAR \citep{dao}.  For each
WFI chip a catalog listing the instrumental $U,~B,~V,$ and $I$ magnitudes was
obtained by cross-correlating the single-band catalogs.  Several hundred
stars in common with \citet{u_cal}, \citet{Sstd}, and \citet{fe92} have been
used to transform the instrumental $U$, $B$, $V$, and $I$ magnitudes to the
Johnson/Cousins photometric system.

As for the WFI data, also for GALEX observations standard photometry and PSF
fitting were performed independently on each image using DAOPHOTII/ALLSTAR.  A
combined FUV-NUV catalog was then obtained by cross-correlating the
single-band catalogs.


\subsection{Astrometry and homogenization of the catalogs }
The HST, WFI, and GALEX catalogs have been placed on the absolute astrometric
system by adopting the procedure already described in \citet{fe01,fe03}.  The
new astrometric Guide Star Catalog (GSC-II\footnote{Available at {\tt
http://www-gsss.stsci.edu/Catalogs/GSC/GSC2/GSC2.htm}.}) was used to search
for astrometric standard stars in the WFI FoV, and a cross-correlation tool
specifically developed at the Bologna Observatory (Montegriffo et al. 2003,
private communication) has been employed to obtain an astrometric solution
for each WFI chip. Several hundred GSC-II reference stars were found in each
chip, thus allowing an accurate absolute positioning of the stars.  Then, we
used more than 3000 and 1500 bright WFI stars in common with the HST and
GALEX samples, respectively, as secondary astrometric standards, so as to
place all the catalogs on the same absolute astrometric system.  We estimate
that the global uncertainties in the astrometric solution is of the order of
$\sim 0\farcs 2$, both in right ascension ($\alpha$) and declination
($\delta$).

Once placed on the same coordinate system, the catalogs have been
cross-correlated and the stars in common have been used to transform all the
magnitudes in the same photometric system. In particular, the HST STMAG
magnitudes have been converted to the WFI ones by using the stars in common
between the two samples in the optical bands.  Then, the GALEX FUV and NUV
instrumental magnitudes have been calibrated onto the HST $m_{160}$ and
$m_{218}$ magnitudes, respectively using the stars in common between
the GALEX and HST samples.

At the end of the procedure a homogeneous master catalog of magnitudes and
absolute coordinates of all the stars included in the HST, WFI, and GALEX
samples was finally produced.

\subsection{Center of gravity and definition of the samples}
\label{sec:cgrav}
Once the absolute positions of individual stars have been obtained, the
center of gravity $C_{\rm grav}$ of NGC~1904 has been determined by averaging
the coordinates $\alpha$ and $\delta$ of all stars lying in the PC FoV,
following the iterative procedure described in \citet[][see also Ferraro et
al. 2003, 2004]{mont95}.  In order to correct for spurious effects due to
incompleteness in the very inner regions of the cluster, we considered two
samples with different limiting magnitudes ($V<19$ and $V<20$), and we
computed the barycenter of stars for each sample.  The two estimates agree
within $\sim 1\arcsec$, setting $C_{\rm grav}$ at $\alpha({\rm J2000}) =
05^{\rm h}\, 24^{\rm m}\, 11\fs 09$, $\delta ({\rm J2000})= -24^{\rm o}\,
31\arcmin\, 29\farcs 00$.
The newly determined center of gravity is located at $\sim 7\arcsec$
south-est ($\Delta\alpha = 7\farcs 3$, $\Delta\delta=-2\arcsec$) from that
previously derived by \citet{har96} on the basis of the surface brightness
distribution.

\begin{figure}[!p]
\begin{center}
\includegraphics[scale=0.7]{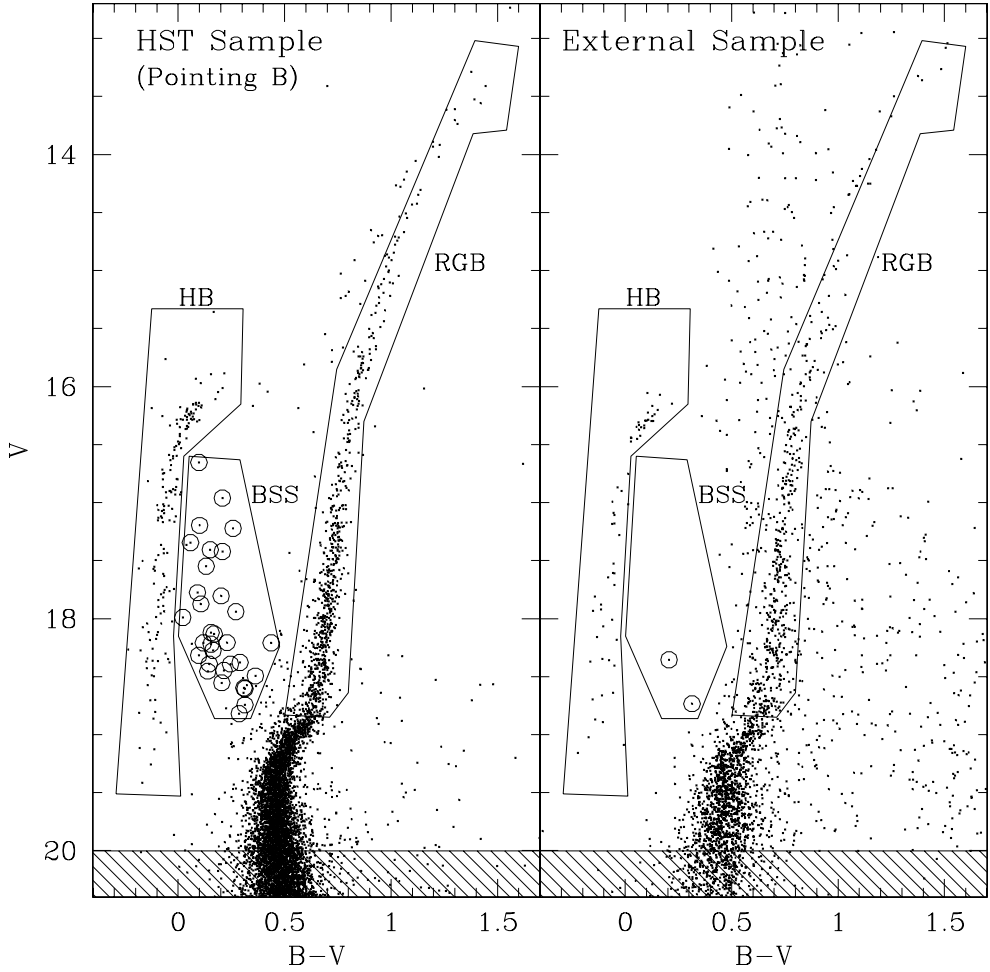}
\caption{($V,~B-V$) CMDs of the HST ({\it Pointing B}) and External samples.
The hatched regions ($V\ge 20$) indicate the stars not used to derive the
cluster surface density profile. The adopted BSS and HB selection boxes are
shown, and all the identified BSS are marked with the empty circles.}
\label{fig:VBV}
\end{center}
\end{figure}

In order to reduce spurious effects in the most crowded regions of the
cluster due to the low resolution of the WFI and GALEX observations, we
considered only the HST data for the inner $85\arcsec$ from the center, this
value being imposed by the geometry of the combined WFPC2 FoVs (see Figure
\ref{fig:HST}).  Thus, in the following we define as \emph{HST sample} the
ensemble of all the stars observed with HST at $r\le 85\arcsec$ from $C_{\rm
grav}$, and as \emph{External sample} all the stars detected with WFI and/or
GALEX at $r>85\arcsec$, out to $\sim 1100\arcsec$ (see Figure \ref{fig:Ext}).
The CMDs of the HST and External samples in the $(V,~B-V)$ planes are shown
in Figure \ref{fig:VBV}.

Note that only the data suitable for the study of the BSS population will be
considered in the following, while those obtained through filters F160BW and
FUV on board HST and GALEX, respectively, will be used in a forthcoming paper
specifically devoted to the analysis the HB properties (Schiavon et
al. 2007).

\subsection{Density profile}
Considering all the stars brighter than $V=20$ in the combined HST+External
catalog (see Figure \ref{fig:VBV}),
we have determined the projected density profile of NGC~1904 by direct star
counts over the entire cluster extension.  Following the procedure already
described in \citet[][2004]{fe99a}, we have divided the entire sample in 31
concentric annuli, each centered on $C_{\rm grav}$ and split in an adequate
number of sub-sectors (quadrants for the annuli totally sampled by the
observations, octants elsewhere). The number of stars lying within each
sub-sector was counted, and the star density was obtained by dividing these
values by the corresponding sub-sector areas.  The stellar density in each
annulus was then obtained as the average of the sub-sector densities, and the
standard deviation was estimated from the variance among the sub-sectors.

\begin{figure}[!p]
\begin{center}
\includegraphics[scale=0.7]{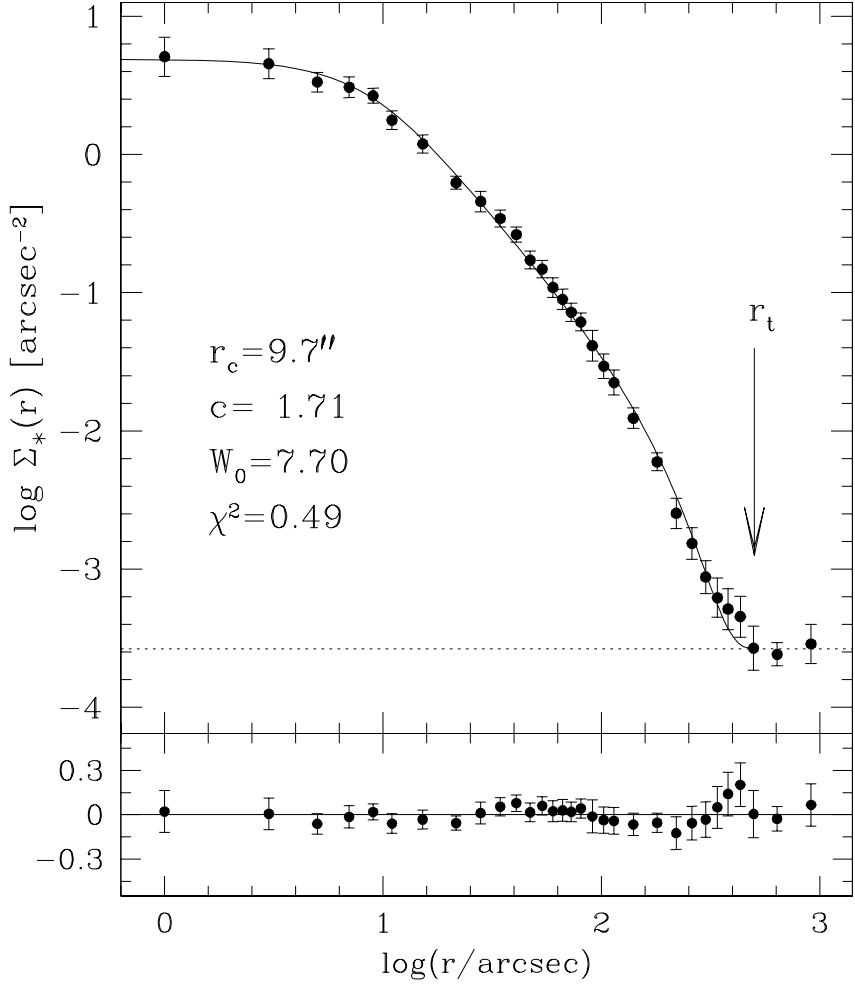}
\caption{Observed surface density profile (dots and error bars) and best-fit
King model (solid line). The radial profile is in units of number of stars
per square arcsec.  The dotted line indicates the adopted level of the
background, and the model characteristic parameters (core radius $r_c$,
concentration $c$, dimensionless central potential $W_0$), as well as the
$\chi^2$ value of the fit are marked in the figure.  The location of the
cluster tidal radius is marked by the arrow. The lower panel shows the
residuals between the observations and the fitted profile at each radial
coordinate.}
\label{fig:prof}
\end{center}
\end{figure}

The radial density profile thus derived is plotted in Figure \ref{fig:prof},
and the average of the three outermost ($r>8\farcm 3$) surface density
measures has been adopted as the background contribution (corresponding to
$0.95$ arcmin$^{-2}$).  Figure \ref{fig:prof} also shows the mono-mass King
model that best fits the derived density profile, with the corresponding
values of the core radius and concentration being $r_c\simeq 9\farcs7$ (with
a typical error of $\sim \pm 2\arcsec$) and $c=1.71$, respectively (hence,
the tidal radius is $r_t\simeq 500\arcsec\simeq 50\, r_c$).  These values are
in good agreement with those quoted by \citet[][$r_c=9\farcs 6$ and
$c=1.72$]{har96}, \citet[][$r_c=9\farcs 55$ and $c=1.72$]{TDK93}, and \citet
[][$r_c=10\farcs 3$ and $c=1.68$]{mcL05}, derived from the surface brightness
profile, and they confirm that NGC~1904 has not yet experienced core
collapse.  By assuming a distance modulus $(m-M)_0=15.63$ \citep[distance
$d\sim 13.37$ kpc,][]{fe99b}, the derived value of $r_c$ corresponds to $\sim
0.65$ pc. By summing the luminosities of stars with $V\le 20$ observed within
$\sim 4\arcsec$, we estimate that the extinction-corrected central surface
brightness of the cluster is $\mu_{V,0}(0)\simeq 16.20$ mag/arcsec$^2$, in
good agreement with \citet[][$\mu_{V,0}=16.23$]{har96},
\citet[][$\mu_{V,0}=16.15$]{djorg93}, and \citet[][$\mu_{V,0}=16.18$]{mcL05}.
Following the procedure described in \citet[][see also Beccari et
al. 2006]{djorg93}, we derive $\log\nu_0 \simeq 3.97$, where $\nu_0$ is the
central luminosity density in units of L$_\odot/$pc$^3$ \citep[for
comparison, $\log\nu_0=4.0$ in][]{har96,djorg93,mcL05}.



\section{THE BSS POPULATION OF NGC~1904 }
\label{sec:samples}
\subsection{BSS selection}
\begin{center}
\begin{figure}[!p]
\includegraphics[scale=0.7]{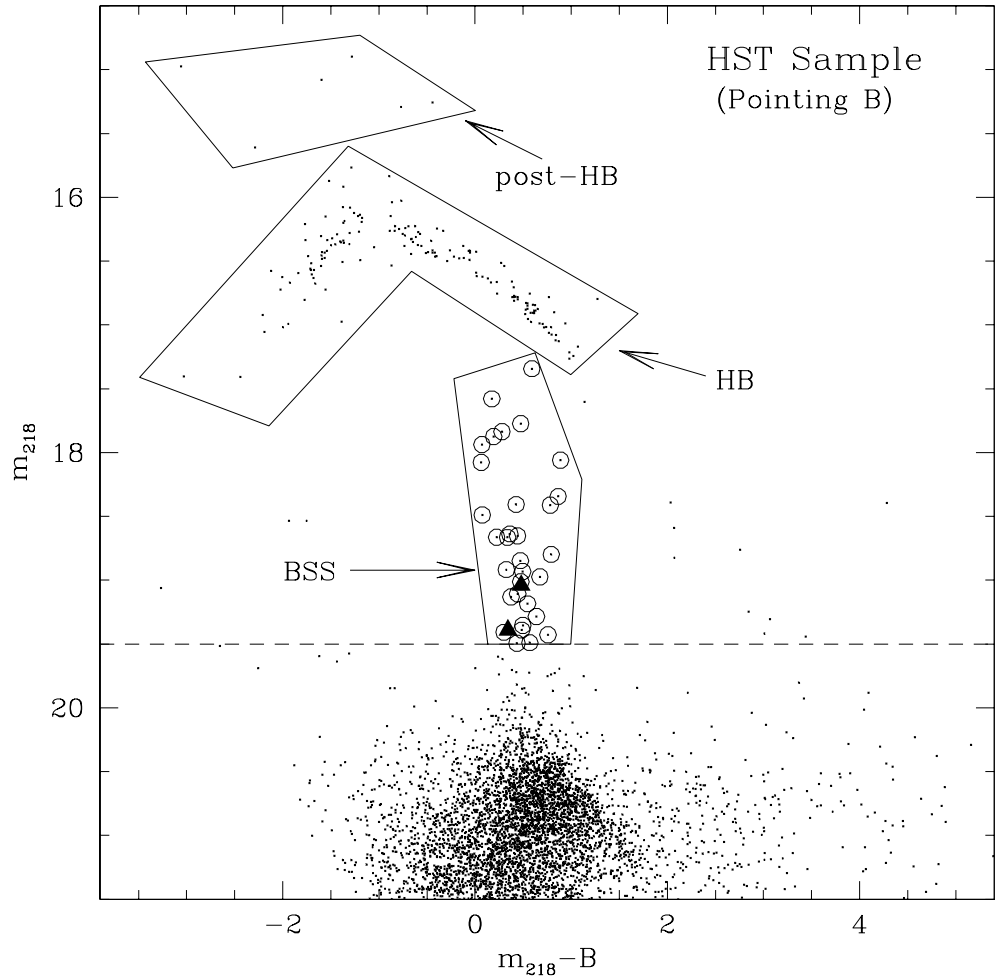}
\caption{CMD of the ultraviolet ({\it Pointing B}) HST sample. The adopted
magnitude limit and selection box used for the definition of the BSS
population (empty circles) are shown.  The two solid triangles correspond to
BSS-38 and 39 found in the External Sample, with UV magnitudes obtained
through the GALEX NUV detector. The selection boxes adopted for HB and
post-HB stars are also shown.}
\label{fig:UVCMD}
\end{figure}
\end{center}

\begin{figure}[!p]
\begin{center}
\includegraphics[scale=0.7]{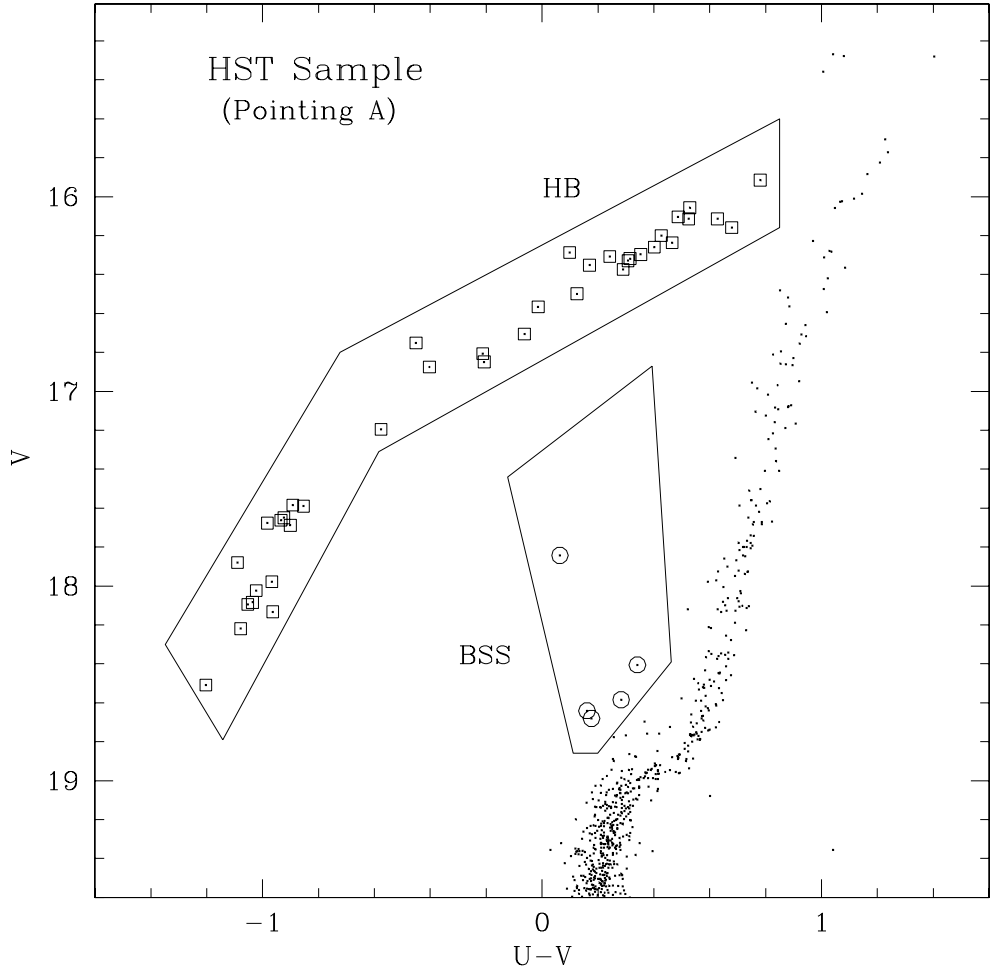}
\caption{($V,~U-V$) CMD of the HST ({\it Pointing A}) sample (only stars not
observed in {\it Pointing B} are plotted).  The adopted BSS and HB selection
boxes are shown, and all the identified BSS and HB stars are marked with the
empty circles and squares, respectively.}
\label{fig:VUV}
\end{center}
\end{figure}

At UV wavelengths BSS are among the brightest objects in a GC, and RGB stars
are particularly faint.  By combining these advantages with the
high-resolution capability of HST, the usual problems associated with
photometric blends and crowding in the high density central regions of GCs
are minimized, and BSS can be most reliably recognized and separated from the
other populations in the UV CMDs.
For these reasons our primary criterion for the definition of the BSS sample
is based on the position of stars in the ($m_{218},~m_{218}-B$) plane
\citep[see also][for a detailed discussion of this issue]{fe04}.
In order to avoid incompleteness bias and the possible contamination from TO
and sub-giant branch stars, we have adopted a limiting magnitude
$m_{218}=19.5$, roughly corresponding to 1 magnitude brighter than the
cluster TO.  The resulting BSS selection box in the UV CMD is shown in Figure
\ref{fig:UVCMD}.
Once selected in the UV CMD, all the BSS lying in the field in common
with the optical-HST sample have been used to define the selection box
in the ($V,~B-V$) and ($V,~U-V$) planes.  The limiting magnitude in
the $V$ band is $V\simeq 18.9$, and the adopted BSS selection boxes in
these planes are shown in Figures \ref{fig:VBV} and \ref{fig:VUV}
(only stars not observed in HST-Pointing B are shown in the latter).

With these criteria we have identified 39 BSS in NGC~1904: 37 in the HST
sample (32 from HST-Pointing B, and 5 from HST-Pointing A) and 2 in the
External sample ($r >85\arcsec$), the most distant lying at $r\simeq
270\arcsec$ ($\sim 4\farcm5$) from the cluster center (see Figure
\ref{fig:Ext}).  All candidate BSS have been confirmed by visual inspection,
evaluating the quality and the precision of the PSF fitting. This procedure
significantly reduces the possibility of introducing spurious objects, such
as blends, background galaxies, etc., in the sample.  The coordinates and
magnitudes of all the identified BSS are listed in Table \ref{tab:BSS}.

In order to study the radial distribution of BSSs, one needs to compare their
number counts as a function of radius with those of a population assumed to
trace the radial density distribution of normal cluster stars.  We chose to
use HB stars for that purpose, given their high luminosities and relatively
large number.  Thanks to the (essentially blue) HB morphology, such a
population can be easily selected in all CMDs, and the adopted selection
boxes, designed to include the bulk of HB and the few post-HB stars, are
shown in Figures \ref{fig:UVCMD}--\ref{fig:VUV}.  In order to be
conservative, a few stars lying within the adopted HB selection boxes in the
optical bands, but not detected in the UV filters (GALEX-NUV channel and
HST-F218W filter), have been excluded from the following analysis.  However
slightly different boxes or the inclusion of these stars in the sample have
no effects on the results.  With these criteria we have identified 249 HB
stars (197 at $r\le 85\arcsec$ from the HST sample, and 52 at $85\arcsec<
r\le r_t$ from the External sample).

\subsection{BSS mass distribution}
\label{sec:BSSmass}
The position of BSS in the CMD can be used to derive a "photometric "
estimate of their masses through the comparison with theoretical isochrones.
We did this in the $(V,~B-V)$ plane, where 34 BSS (32 from the HST-Pointing B
and 2 from the External sample) out of the 39 identified in the cluster have
been measured.

\begin{figure}[!p]
\begin{center}
\includegraphics[scale=0.7]{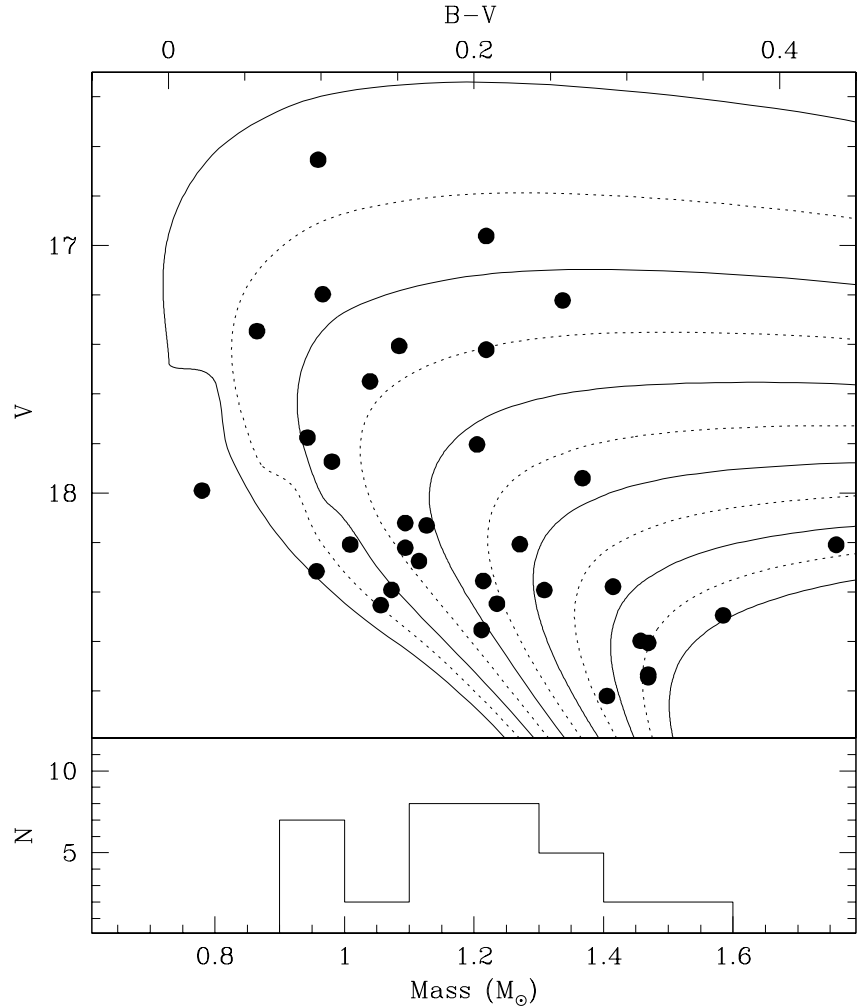}
\caption{{\it Upper panel:} zoomed ($V,~B-V$) CMD of the BSS region; the 34
BSS measured in this plane are shown. The set of isochrones ranging from 1 to
6 Gyr (stepped by 0.5 Gyr) from \citet{cari03} data base used to derive BSS
masses is also shown.  {\it Lower panel:} derived mass distribution for the
BSS shown in the upper panel.}
\label{fig:BSSmass}
\end{center}
\end{figure}

A set of isochrones of appropriate metallicity ($Z=6\times 10^{-4}$) has been
extracted from the data-base of \citet{cari03} and transformed into the
observational plane by adopting a reddening $E(B-V)=0.01$ \citep{fe99b}. The
12 Gyr isochrone nicely reproduces the main cluster population, while the
region of the CMD populated by the BSS is well spanned by a set of isochrones
with ages ranging from 1 to 6 Gyr (see Figure \ref{fig:BSSmass}).  Thus, the
entire dataset of isochrones available in this age range (stepped at 0.5 Gyr)
has been used to derive a grid linking the BSS colors and magnitudes to their
masses. Each BSS has been projected on the closest isochrone and a value of
its mass has been derived.  As shown in the lower panel of Figure
\ref{fig:BSSmass}, BSS masses range from $\sim 0.95$ to $\sim 1.6 M_\odot$,
and both the mean and the median of distribution correspond to 1.2
$M_\odot$. The TO mass turns out to be $M_{TO}=0.8 M_\odot$.

\subsection{The BSS radial distribution}
\label{sec:radist}
\begin{figure}[!p]
\begin{center}
\includegraphics[scale=0.7]{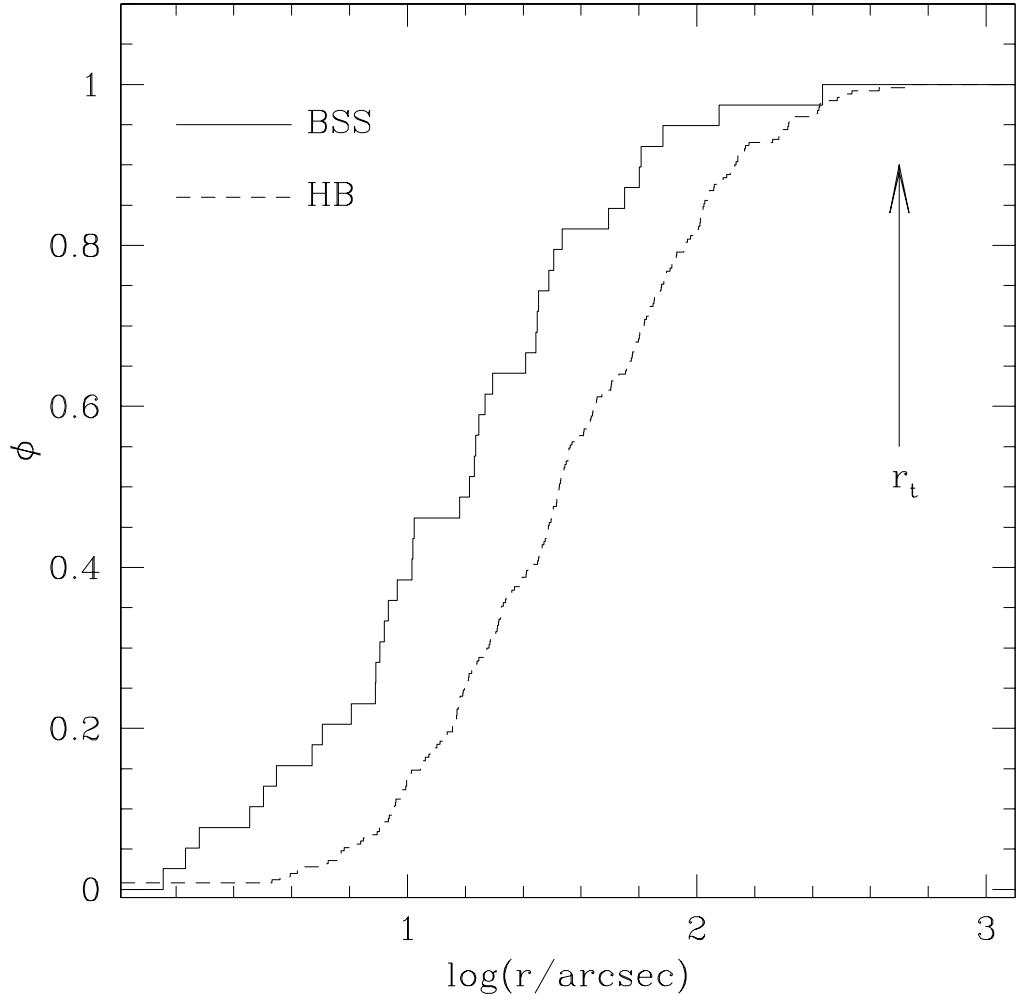}
\caption{Cumulative radial distribution of BSS (solid line) and HB (dashed
line) stars as a function of the projected distance from the cluster center
for the combined HST+External sample. The location of the cluster tidal
radius is marked by the arrow.}
\label{fig:KS}
\end{center}
\end{figure}

The radial distribution of BSS identified in NGC~1904 has been studied
following the same procedure previously adopted for other clusters \citep[see
references in][]{onlyfe06, bec06}.  In Figure \ref{fig:KS} we compare the BSS
cumulative radial distribution to that of HB stars.  The two distributions
are obviously different, with the BSS being more centrally concentrated than
HB stars.  A Kolmogorov-Smirnov test gives a $\sim 7 \times 10^{-4}$
probability that they are extracted from the same population, i.e. the two
populations are different at more than $3\sigma$ level.

For a more quantitative analysis, the surveyed area has been divided into 6
concentric annuli, the first roughly corresponding to the core radius
($r=10\arcsec$), and the others chosen in order to sample approximately the
same fraction of the cluster luminosity out to the tidal radius ($r_t\simeq
500\arcsec$).  The luminosity in each annulus has been calculated by
integrating the surface density profile shown in Figure \ref{fig:prof}.  
The number of BSS and HB stars ($N_{\rm BSS}$ and $N_{\rm HB}$,
respectively), as well as the fraction of sampled luminosity ($L^{\rm samp}$)
measured in each annulus are listed in Table \ref{tab:annuli} and have been
used to compute the population ratio $N_{\rm BSS}/N_{\rm HB}$ and the
specific frequencies \citep[see][]{fe03}:
\begin{equation}
R_{\rm pop}=\frac{(N_{\rm pop}/N_{\rm pop}^{\rm tot})}{(L^{\rm samp}/L_{\rm tot}^{\rm samp})}, 
\label{eq:spec_freq}
\end{equation}
with pop = BSS, HB.

The resulting radial trend of $R_{\rm HB}$ over the surveyed area is
essentially constant, with a value close to unity (see Figure
\ref{fig:Rpop}). This is just what expected on the basis of the stellar
evolution theory, which predicts that the fraction of stars in any post-main
sequence evolutionary stage is strictly proportional to the fraction of the
sampled luminosity \citep{renffp88}.  In contrast the BSS show a completely
different radial distribution: as shown in Figure \ref{fig:Rpop}, the
specific frequency $R_{\rm BSS}$ is highly peaked at the cluster center
decreases to a minimum at $r\simeq 12\, r_c$ and remains approximately
constant outwards.  The same behavior is clearly visible also in Figure
\ref{fig:simu}, where the population ratio $N_{\rm BSS}/N_{\rm HB}$ is
plotted as a function of $r/r_c$.

\subsection{Dynamical simulations}
\label{sec:simu}

Following the same approach as \citet[][2006]{ma04} and \citet{lan07}, we
have used a Monte-Carlo simulation code \citep[originally developed
by][]{sigu95} in order to reproduce the observed radial distribution and to
derive some clues about the BSS formation mechanisms.  Such a code follows
the dynamical evolution of $N$ BSS within a background cluster, taking into
account the effects of both dynamical friction and distant encounters.  Since
stellar collisions are most probable in the central high-density regions of
the clusters, in the simulations we define COL-BSS those objects with initial
positions $r_i \lsim r_c$.  Since primordial binaries most likely evolve in
isolation if they orbit in the cluster outskirts, we identify as MT-BSS those
BSS having $r_i \gg r_c$.  Within these defintions, in any given run we
assume that a certain fraction of the $N$ simulated BSS is made of COL-BSS
and the remaining fraction of MT-BSS.  The initial positions $r_i$ of the two
types of BSS are randomly generated within the appropriate radial range ($r_i
\lsim r_c$ for COL-BSS, and $r_i \gg r_c$ for the others) following a flat
distribution, according to the fact that the number of stars in a King model
scales as $dN = n(r)\,dV \propto r^{-2}\pi r^2 dr\propto dr$.  Their initial
velocities are randomly extracted from the cluster velocity distribution
illustrated in \citet{sigu95}, and an additional natal kick is assigned to
COL-BSS to account for the recoil induced by the three-body encounters that
trigger the merger and produce the BSS \citep[see, e.g.,][]{sigu94, dav94}.
Each BSS has characteristic mass $M$ and maximum lifetime $t_{\rm last}$. We
follow their dynamical evolution in the (fixed) gravitational potential for a
time $t_i$ ($i=1,N$), where each $t_i$ is a randomly chosen fraction of
$t_{\rm last}$.  At the end of the simulation we register the final positions
of BSS, and we compare their radial distribution with the observed one. The
percentage of COL- and MT-BSS is changed and the procedure repeated until a
reasonable agreement between the simulated and the observed distributions is
reached.

For a more detailed discussion of the procedure and the ranges of values
appropriate for the input parameters we refer to \citet{ma06}.  Here we only
list the assumptions made in the present study:

\begin{itemize}
\item[--]the background cluster has been approximated with a
multi-mass King model, determined as the best fit to the observed
profile\footnote{By adopting the same mass groups as those of
\citet{ma06}, the resulting value of the King dimensionless central
potential is $W_0=10$}. The cluster central velocity dispersion is set
to $\sigma= 3.9\,{\rm km\,s^{-1}}$ \citep{dub97}, and, assuming $0.5\,
M_\odot$ as the average mass of the cluster stars, the central stellar
density is $n_c=3\times 10^4\,{\rm pc^{-3}}$ \citep{pry93};

\item[--]BSS masses have been fixed to $M=1.2\, M_\odot$ (see Section
\ref{sec:BSSmass}) and characteristic lifetimes $t_{last}$ ranging between
1.5 and 4 Gyr have been considered;
\item[--]COL-BSS have been distributed with initial positions $r_i\le r_c$
and have been given a natal kick velocity of $1\times\sigma$;
\item[--]initial positions ranging between $5\,r_c$ and $r_t$ have been
considered for MT-BSS in different runs;
\item[--]in each simulation we have followed the evolution of $N=10,000$ BSS.
\end{itemize}

\begin{figure}[!p]
\begin{center}
\includegraphics[scale=0.8]{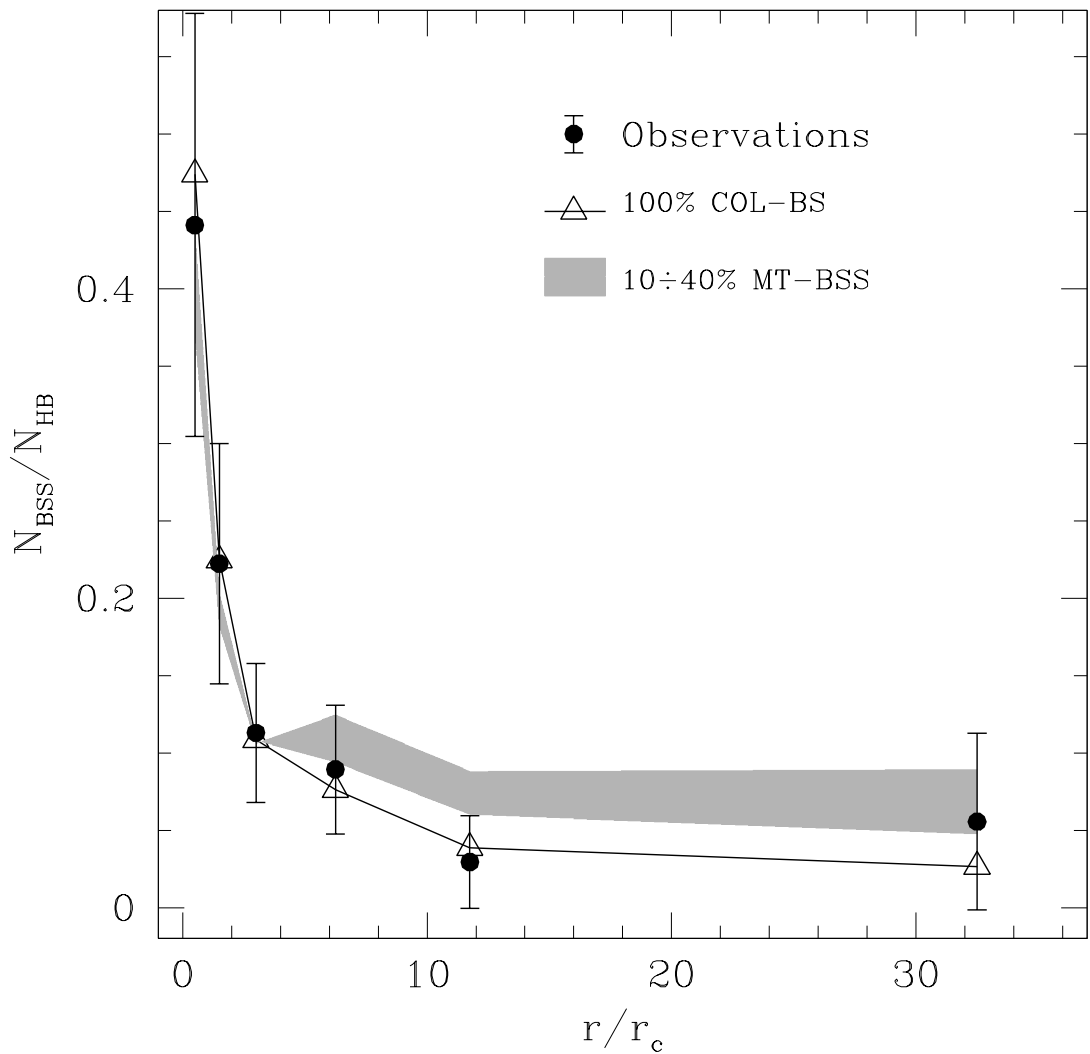}
\caption{Radial distribution of the population ratio $N_{\rm BSS}/N_{\rm HB}$
as a function of $r/r_c$ (dots with error bars), compared with the simulated
distribution (solid line and triangles) obtained by assuming $100\%$ of
COL-BSS.  The results of the simulations obtained by assuming a percentage of
MT-BSS ranging from 10\% to 40\% (lower and upper boundaries of the gray
region, respectively) are also shown.}
\label{fig:simu}
\end{center}
\end{figure}

The simulated radial distribution that best reproduces the observed
one (with a reduced $\chi^2\simeq 0.1$) is shown in
Figure~\ref{fig:simu} and is obtained by assuming that the totality of
BSS is made of COL-BSS.  In the best-fit case the BSS characteristic
lifetime is $t_{\rm last}\simeq 1.5$ Gyr, but a variation between 1
and 4 Gyr of this parameter still leads to a very good agreement
($\chi^2\simeq 0.2$--0.3) with the observations.  For the sake of
comparison, in Figure~\ref{fig:simu} we also show the results of the
simulations obtained by assuming a percentage of MT-BSS ranging from
10\% to 40\% (see lower and upper boundaries of the gray region,
respectively)\footnote{Note that a population of 40\% MT-BSS was
needed in order to reproduce the bimodal distribution observed in M~3,
47~Tuc and NGC~6752 \citep{ma06}, and 10\% was found to be the
appropriate percentage of MT-BSS in the case of M~5 \citep{lan07}.}.
As can be seen, while a population of 10\% MT-BSS is still marginally
consistent with the observations, larger percentages systematically
overestimate the BSS population at $r\gsim 5\,r_c$.  Increasing the
BSS mass up to $1.5\,M_\odot$ does not change this conclusion.

By assuming 12 Gyr for the age of NGC~1904, we have used the simulations and
the dynamical friction timescale \citep[from, e.g.,][]{ma06} for $ 1.2\,
M_\odot$ stars to estimate the radius of avoidance $r_{\rm avoid}$ of the
cluster, i.e., the radius within which all these stars are expected to have
already sunk to the cluster core because of mass segregation processes.  We
find that $r_{\rm avoid}\sim 30\,r_c$ (i.e., $\sim 300\arcsec$), which
corresponds to a significant fraction of the entire cluster extension. This
evidence is consistent with the fact that the simulated MT-BSS appear to be a
negligible fraction of the overall BSS population.

\section{DISCUSSION}
\begin{figure}[!p]
\begin{center}
\includegraphics[scale=0.7]{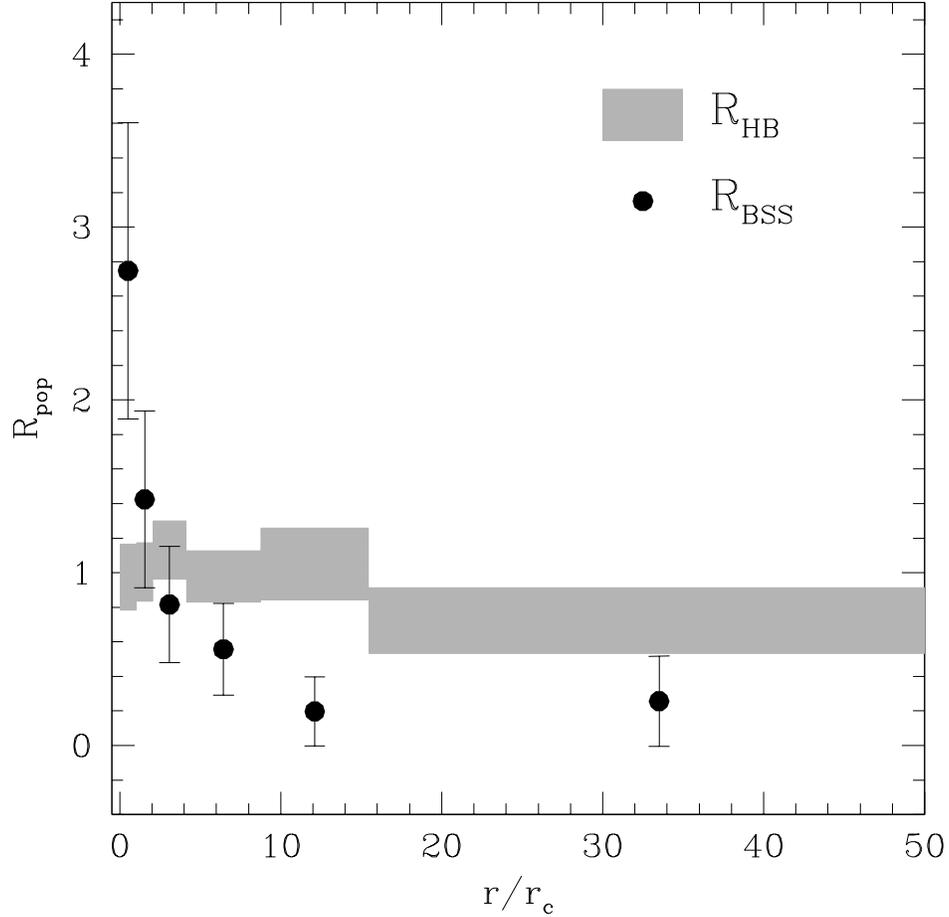}
\caption{Radial distribution of the BSS (dots) and HB (gray regions) specific
frequencies, as defined in equation~(\ref{eq:spec_freq}), and as a function
of the radial distance in units of the core radius. The vertical size of the
gray regions correspond to the error bars.}
\label{fig:Rpop}
\end{center}
\end{figure}

We have studied the brightest portion ($m_{218}\le19.5$) of the BSS
population in NGC~1904. We have found a total of 39 objects, with a
high degree of segregation in the cluster center. Approximately 38\%
of the entire BSS population is found within the cluster core, while
only $\sim 13\%$ of HB stars are counted in the same region. This
indicates a significant overabundance of BSS in the center, as also
confirmed by the fact that the BSS specific frequency $R_{\rm BSS}$
within $r_c$ is roughly 3 times larger than expected for a normal
(non-segregated) population on the basis of the sampled light (see
Figure \ref{fig:Rpop}).  The peak value is in good agreement with what
is found in the case of M~3, 47~Tuc, NGC~6752 and M~5
\citep[see][]{fe04,sab04,lan07}.  Unlike these clusters, no
significant upturn of the distribution at large radii has been
detected in NGC~1904.  

\begin{figure}[!p]
\begin{center}
\includegraphics[scale=0.8]{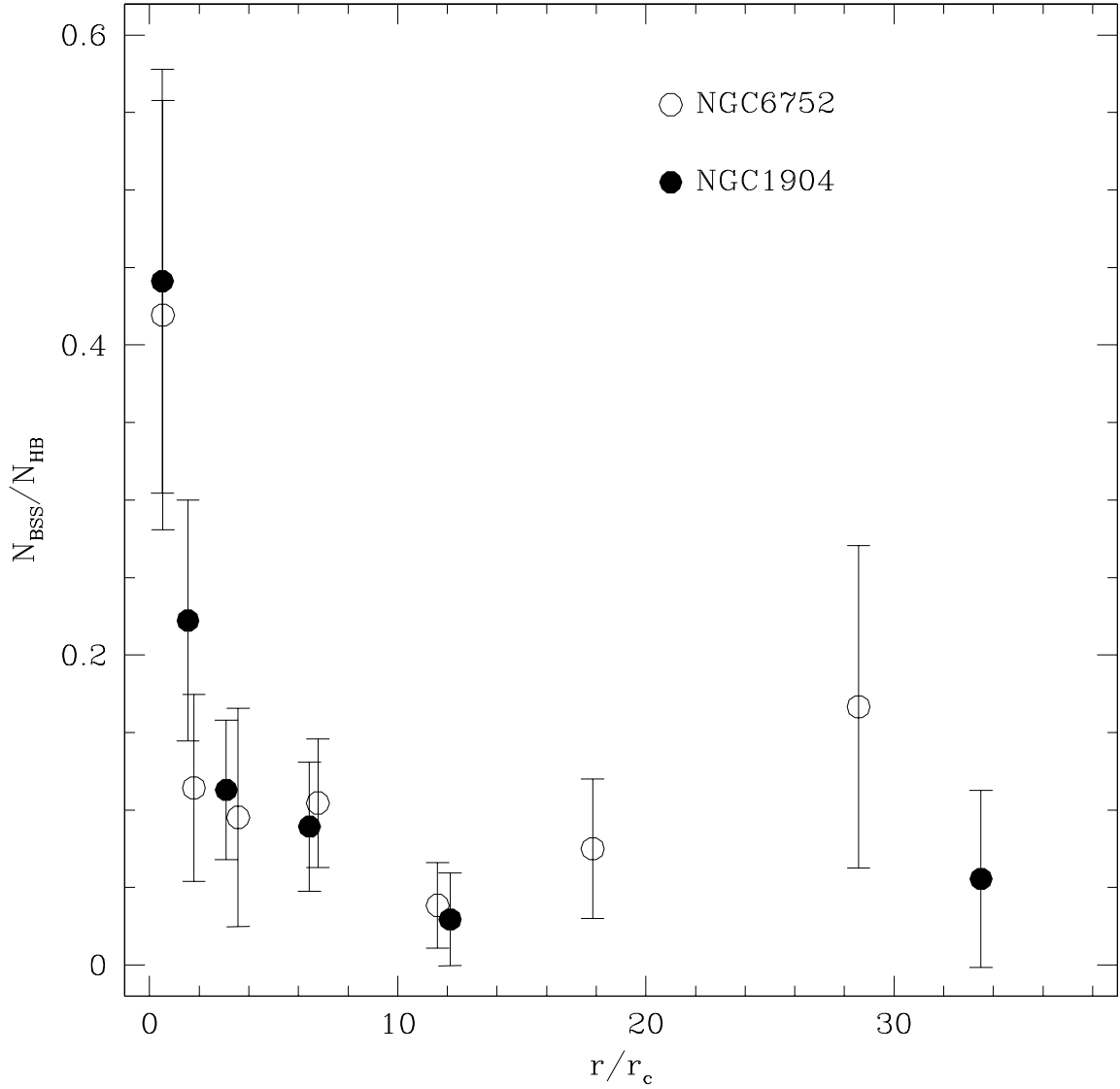}
\caption{Radial distribution of the population ratio $N_{\rm BSS}/N_{\rm HB}$
for NGC~1904 {\it (filled circles)} and NGC~6752 {\it (open circles)} plotted
as a function of the radial distance in core radius units.}
\label{fig:N6752}
\end{center}
\end{figure}

We emphasize that the absence of an external upturn in the BSS radial
distribution is not an effect of low statistics. In the case of NGC~6752,
where a similar amount of BSS (34) has been detected, the BSS radial
distribution is clearly bimodal \citep{sab04}. This can be seen also in
Figure \ref{fig:N6752}, where the two distributions are directly
compared. They nicely agree within $r\sim 12 r_c$, but the fraction of BSS in
NGC~6752 rises again at larger distances from the center, 
despite the smaller number of BSS observed in this cluster compared to
NGC~1904.

Extensive dynamical simulations have been used to derive some hints about the
BSS formation mechanisms. Even if admittedly crude, this approach has been
successfully used to demonstrate that the external rising branch of the BSS
radial distribution observed in M~3, 47~Tuc, NGC~6752 and M~5 cannot be due
to COL-BSS originated in the core and then kicked out in the outer regions:
hence, a significant fraction (20-40\%) of the overall population is required
to be made of MT-BSS in these clusters \citep{ma06, lan07}.  By using the
same simulations to interpret the (flat) BSS radial distribution of NGC~1904,
we found that only a negligible percentage (0--10\%) of MT-BSS is
needed. However, we emphasize that if a rising peripheral BSS frequency is
absent (as in the case of NGC~1904) our simple approach cannot distinguish
between BSS created by MT (and then segregated into the cluster core by the
dynamical friction) and COL-BSS created by collisions inside the core.

On the other hand, the negligible fraction of peripheral MT-BSS found in
NGC~1904 is in agreement with the quite large value of the radius of
avoidance estimated for this cluster ($r_{\rm avoid}\simeq 30\,r_c$), which
indicates that all the heavy stars (binaries) within this radial distance
have had enough time to sink to the core and are therefore not expected in
the cluster outskirts. Such a radial distance corresponds to $0.6\,r_t$,
i.e., it represents a significant fraction of the cluster extension (only 1\%
of the cluster light is contained between $r_{\rm avoid}$ and $r_t$), and
hence only a small fraction of the massive objects are expected to be
unaffected by the dynamical friction).  In all the other studied cases,
$r_{\rm avoid}$ is significantly smaller: $r_{\rm avoid}\lsim 0.2\,r_t$
\citep{ma06, lan07}.  In turn, this suggests that at least a fraction of the
BSS population that we now observe in the cluster center are primordial
binaries which have sunk to the core because of the dynamical friction
process, and mixed with those that formed through stellar collisions.

Only systematic surveys of physical and chemical properties for a large
number of BSS in different environments \citep[see examples in][]{demarco05,
COdep} can definitively identify the formation processes of these stars.




\acknowledgements{This research was supported by Agenzia Spaziale Italiana
under contract ASI-INAF I/023/05/0, by the Istituto Nazionale di Astrofisica
under contract PRIN/INAF 2006, and by the Ministero dell'Istruzione,
dell'Universit\`a e della Ricerca. RTR is partially funded by NASA through
grant number HST-GO-10524 from the Space Telescope Science Institute.}

\begin{deluxetable}{lccccccc}
\footnotesize
\tablewidth{14.5cm}
\tablecaption{The BSS population in NGC1904}
\startdata \\
\hline \hline
Name   &    RA       &    DEC      &$m_{218}$& U   &   B   &   V   &  I   \\
       & [degree]    &   [degree]  &       &       &       &       &      \\
\hline
BSS-1  & 81.048797100 & -24.526391100 & 19.11 & 18.64 & 18.66 & 18.45 &  -  \\ 
BSS-2  & 81.047782800 & -24.526527400 & 18.65 & 18.12 & 18.21 & 17.94 &  -  \\ 
BSS-3  & 81.047540400 & -24.526005300 & 18.85 & 18.34 & 18.38 & 18.22 &  -  \\ 
BSS-4  & 81.048954000 & -24.525188200 & 17.94 & 17.75 & 17.87 & 17.78 &  -  \\ 
BSS-5  & 81.047199200 & -24.525797600 & 17.58 & 17.47 & 17.40 & 17.35 &  -  \\ 
BSS-6  & 81.045528300 & -24.525876800 & 18.64 & 18.24 & 18.28 & 18.12 &  -  \\ 
BSS-7  & 81.044296100 & -24.526362200 & 19.13 & 18.77 & 18.76 & 18.55 &  -  \\ 
BSS-8  & 81.048506700 & -24.524266800 & 19.18 & 18.55 & 18.64 & 18.39 &  -  \\ 
BSS-9  & 81.041556100 & -24.526972400 & 19.35 & 18.65 & 18.86 & 18.49 &  -  \\ 
BSS-10 & 81.045827300 & -24.525062700 & 18.06 & 17.35 & 17.17 & 16.96 &  -  \\ 
BSS-11 & 81.045467700 & -24.525147900 & 17.84 & 17.44 & 17.56 & 17.41 &  -  \\ 
BSS-12 & 81.047088300 & -24.524375900 & 18.80 & 18.20 & 18.01 & 17.80 &  -  \\ 
BSS-13 & 81.046548400 & -24.524483700 & 19.43 & 18.77 & 18.67 & 18.38 &  -  \\ 
BSS-14 & 81.045121300 & -24.524748300 & 19.28 & 18.33 & 18.65 & 18.21 &  -  \\ 
BSS-15 & 81.045883300 & -24.524278600 & 18.97 & 18.42 & 18.30 & 18.13 &  -  \\ 
BSS-16 & 81.046164500 & -24.522567300 & 18.41 & 17.72 & 17.63 & 17.42 &  -  \\ 
BSS-17 & 81.044313700 & -24.523304200 & 18.66 & 18.46 & 18.44 & 18.27 &  -  \\ 
BSS-18 & 81.047157500 & -24.521982700 & 18.49 & 18.45 & 18.41 & 18.32 &  -  \\ 
BSS-19 & 81.043418100 & -24.523247200 & 17.88 & 18.16 & 17.68 & 17.55 &  -  \\ 
BSS-20 & 81.046665900 & -24.520192000 & 19.49 & 18.71 & 18.92 & 18.60 &  -  \\ 
BSS-21 & 81.044991400 & -24.520127500 & 18.93 & 18.45 & 18.44 & 18.21 &  -  \\ 
BSS-22 & 81.046157800 & -24.519245100 & 19.49 & 18.92 & 19.06 & 18.74 &  -  \\ 
BSS-23 & 81.049326100 & -24.521621900 & 18.08 &   -   & 18.01 & 17.99 &  -  \\ 
BSS-24 & 81.049155900 & -24.520629500 & 18.41 &   -   & 17.98 & 17.87 &  -  \\ 
BSS-25 & 81.047244500 & -24.517052600 & 19.41 & 18.97 & 19.11 & 18.82 &  -  \\ 
BSS-26 & 81.051592200 & -24.523146600 & 18.67 &   -   & 18.33 & 18.21 &  -  \\ 
BSS-27 & 81.050476100 & -24.517107000 & 19.01 & 18.65 & 18.54 & 18.39 &  -  \\ 
BSS-28 & 81.060767000 & -24.520983900 & 18.92 &   -   & 18.59 & 18.45 &  -  \\ 
BSS-29 & 81.068117100 & -24.517558600 & 19.39 &   -   & 18.91 & 18.60 &  -  \\ 
BSS-30 & 81.043233100 & -24.533852000 & 17.34 &   -   & 16.75 & 16.65 &  -  \\ 
BSS-31 & 81.044917900 & -24.540315000 & 18.34 &   -   & 17.48 & 17.22 &  -  \\ 
BSS-32 & 81.038520800 & -24.540891600 & 17.77 & 17.46 & 17.30 & 17.20 &  -  \\ 
BSS-33 & 81.045196533 & -24.515874863 &   -   & 17.91 &   -   & 17.84 &  -  \\ 
BSS-34 & 81.032196045 & -24.512256622 &   -   & 18.75 &   -   & 18.41 &  -  \\ 
BSS-35 & 81.037574768 & -24.525493622 &   -   & 18.87 &   -   & 18.58 &  -  \\ 
BSS-36 & 81.041069031 & -24.518457413 &   -   & 18.80 &   -   & 18.64 &  -  \\ 
BSS-37 & 81.045227051 & -24.517648697 &   -   & 18.86 &   -   & 18.68 &  -  \\ 
BSS-38 & 81.056510925 & -24.449676514 & 19.04$^\dag$ & 18.78 & 18.56 & 18.35 & 18.08 \\ 
BSS-39 & 81.058883667 & -24.555763245 & 19.39$^\dag$ & 19.14 & 19.05 & 18.73 & 18.34 \\ 
\hline 
\enddata
\tablecomments{$^\dag$~Note that, while the header of the column referes to
HST-F218W magnitudes, those of BSS-38 and -39 have been obtained with the NUV 
channel of GALEX and transformed to the $m_{218}$ scale as described in
Section 2.2.}
\label{tab:BSS}
\end{deluxetable}
\normalsize

\begin{table}[!hp]
\begin{center}
\begin{tabular}{rrrrc}
\hline \hline
$r_i$ & $r_e$ & $N_{\rm BSS}$ & $N_{\rm HB}$ & $L^{\rm samp}/L_{\rm tot}^{\rm samp}$\\
\hline \hline
  0 &   10 &  15 & 34 &  0.14 \\
  10 &  20 &  10 & 45 &  0.18 \\
  20 &  40 &   7 & 62 &  0.22 \\
  40 &  85 &   5 & 56 &  0.23 \\
  85 & 150 &   1 & 34 &  0.13 \\
 150 & 500 &   1 & 18 &  0.10 \\
\hline \hline
\end{tabular}
\caption{Number of BSS and HB stars, and fraction of luminosity sampled in
the 6 concentric annuli used to study the BSS radial distribution of NGC~1904
($r_i$ and $r_e$ correspond to the internal and external radius of each
considered annulus, in arcsec).}
\label{tab:annuli}
\end{center}
\end{table}

\end{document}